# AN INTRA-PULSE FEEDFORWARD ALGORITHM FOR IMPROVING PULSED MICROWAVE STABILITY*

J.W. Han, H.L. Ding[1†], J.F. Zhu[‡], H.K. Li[1], X.W. Dai, J.Y. Yang[1], W.Q. Zhang[1]
Institute of Advanced Science Facilities, Shenzhen, China
[1]also at Dalian Institute of Chemical Physics, Chinese Academy of Sciences, Dalian, China

*Abstract*

During the pulsed operation of the linear accelerator in DCLS (Dalian Coherent Light Source), we found a strong correlation between the klystron modulator's high voltage and the klystron output microwave, with noticeable jitter among adjacent microwaves. Therefore, we propose an intra-pulse feedforward algorithm and implement it in LLRF (Low-Level Radiofrequency) systems. This algorithm assumes that the transfer model of the microwave system is linear within a small range of work points and measures the transfer coefficient of the microwave between the LLRF and klystron. For each pulsed microwave of the klystron output, the LLRF system first calculates the vector deviation between the initial measurement within its pulse and the target. The deviation will be compensated in the LLRF excitation so that the jitter in the later part of the pulsed microwave is suppressed. Experiments have shown that this algorithm can effectively suppress the jitter among adjacent microwaves, e.g., improving the amplitude and phase stability (RMS) from 0.11%/0.2° to 0.1%/0.05°. This algorithm can also be applied to other accelerators operating in pulsed modes.

## INTRODUCTION

The Free Electron Laser (FEL) device is a fourth-generation light source device with extremely high peak brightness, ultra-short pulses, and fully coherent characteristics[1]. The free electron laser devices built and under construction (including planning) in China include the Dalian Coherent Light Source[2], Shanghai Soft X-ray Free Electron Laser Device[3] and Shenzhen SRF Soft Free Electron Laser (S3FEL). Among them, S3FEL is planned to be an advanced high repetition rate free electron laser device that will cover the entire soft X-ray band. After completion, it will provide cutting-edge research methods for disciplines such as materials science, biomedicine, and physical chemistry.

In the pulse operation process of the Dalian Coherent Light Source (DCLS) linear accelerator, the klystron plays an important role in the microwave system of the linear accelerator. After the low-level control system modulates the amplitude and phase of the input reference signal, the modulated reference signal is first input to the solid-state power amplifier for first amplification, and then the output signal is sent to the high-power klystron for second amplification. After secondary amplification, the output signal of the klystron is sent to the coupler through the transmission waveguide, and finally fed into the accelerator cavity structure, forming a stable and reliable acceleration field to obtain high-quality electron beam pulse trains that meet the requirements. We found a strong correlation between the high voltage of the modulator and the output microwave, and observed significant jitter between adjacent microwaves. In recent years, the methods for improving the amplitude and phase stability of low-level microwave excitation mainly include feedforward control and feedback control. Feedback control mainly suppresses thermal drift and low-frequency noise through PI feedback between pulses [4], which can not solve the amplitude and phase jitter of adjacent microwaves in short pulses. The RF feed-forward can flat RF pulse amplitude and phase of the linear accelerator to provide high beam quality. In this regard, this article mathematically models the physical model of klystron jitter and proposes an intra pulse feedforward algorithm based on the klystron nonlinear amplitude phase jump model. The microwave transfer coefficient between LLRF and klystron is calculated in the software, and real-time compensation is made in the firmware based on the correction matrix to suppress pulse microwave jitter.

## ALGORITHM AND SCHEME DESIGN

### klystron small-scale linear and amplitude phase jump model

The klystron amplifier is a nonlinear device [5], and the abstract expression of the klystron model helps us better solve the klystron amplitude phase jump. The known klystron normalized input and output curve is shown in Figure 1. The abstract expression of the klystron model helps us better solve the klystron amplitude phase jump. Given the normalized input output curve of klystron as shown in Figure 1, we define *kgain* as the small range linear gain of klystron input output. Assuming it is linear in the small range near the operating point $(A_{in(0)}, A_{out(0)})$, then:

$$kgain = \frac{A_{out(i)} - A_{out(0)}}{A_{in(i)} - A_{in(0)}} = \frac{\Delta A_{out(i)}}{\Delta A_{in(i)}} \tag{1}$$

* Work supported by the National Natural Science Foundation of China (Grant No. 22288201), the Scientific Instrument Developing Project of the Chinese Academy of Sciences (Grant No. GJJSTD20220001), and the Shenzhen Science and Technology Program (Grant No. RCBS20221008093247072).
† dinghongli@dicp.ac.cn
‡ zhujinfu@mail.iasf.ac.cn

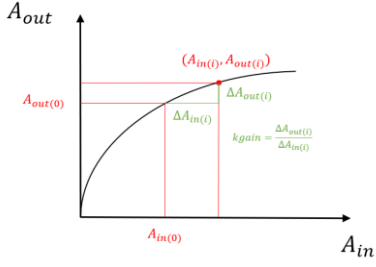

Figure 1: klystron Normalized Input Output Curve

The klystron jump can cause the amplitude and phase of the klystron output signal to be inconsistent with the target value set by the user. The mathematical expression for klystron amplitude and phase jump is: the target value set by the user is $(Acos\varphi, Asin\varphi)$. After the klystron amplitude phase jump, this value is $((A + \Delta A)cos(\varphi + \Delta\varphi), (A + \Delta A)sin(\varphi + \Delta\varphi))$. In practical engineering, the measurement of klystron amplitude phase jump output value by users will introduce loop phase shift and attenuation. Based on this, as shown in Figure 2, a unified mathematical model for klystron small range amplitude phase jump, small range approximate linear gain, and loop measurement is established. In the model, $sgain$ is defined as the gain from the user excitation amplitude set value to the klystron output amplitude measurement value; $\theta$ is the phase shift from the user excitation phase setting value to the klystron output phase measurement value.

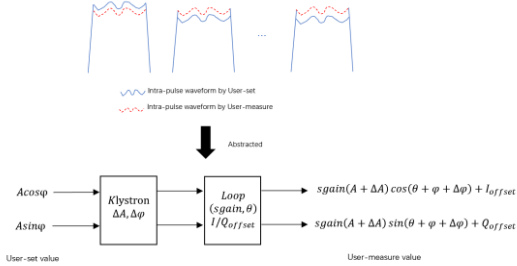

Figure 2: klystron small-scale linear and amplitude phase jump model

### Intra-pulse feedforward algorithm

To solve the problem of klystron amplitude and phase jump in a small-scale linear amplitude and phase jump model, we designed an intra pulse feedforward algorithm. The main idea of the algorithm is to make a difference between the amplitude and phase values after the amplitude and phase jump and the target value, bring the difference into the inverse transformation of the klystron amplitude and phase jump, and then obtain the compensation value of the klystron input to compensate for the user's set value. So, after compensating for the user set value and undergoing klystron amplitude phase jump, we can obtain the ideal target value output. The mathematical reasoning process of the algorithm is shown in Figure 3:

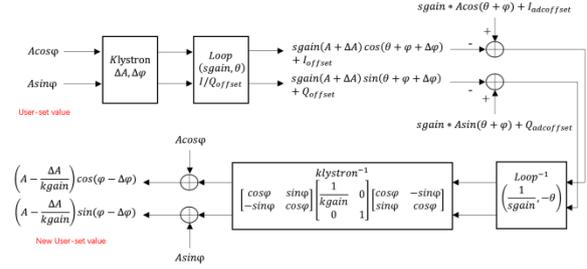

Figure 3: Mathematical derivation of intra pulse feedforward algorithm

The target value set by the known user is $(Acos\varphi, Asin\varphi)$, the measured value after the klystron jump is $(sgain(A + \Delta A)cos(\theta + \varphi + \Delta\varphi) + I_{adcoffset}, sgain(A + \Delta A)sin(\theta + \varphi + \Delta\varphi) + Q_{adcoffset})$. Our goal is to eliminate the amplitude phase jump of klystron and obtain the target value set by the user at the klystron output end.

Firstly, based on the amplitude phase jump property, averaging multiple measurements can eliminate the klystron jump. Based on this, the ideal measurement value for klystron without jump is $(sgain * Acos(\theta + \varphi) + I_{adcoffset}, sgain * Asin(\theta + \varphi) + Q_{adcoffset})$. The difference between this value and the measured value after the jump of the klystron is $(sgain * Acos(\varphi + \theta) - sgain * (A + \Delta A)cos(\theta + \varphi + \Delta\varphi), sgain * Asin(\varphi + \theta) - sgain * (A + \Delta A)sin(\theta + \varphi + \Delta\varphi))$.

Then, by performing the inverse transformation of loop phase shift and attenuation on the difference, the error term $(Acos\varphi - (A + \Delta A)cos(\varphi + \Delta\varphi), Asin\varphi - (A + \Delta A)sin(\varphi + \Delta\varphi))$ at the klystron output end can be obtained.

Finally, based on the nonlinear gain of klystron, it can be inferred that to convert the error term at the klystron output end to the input end and compensate for it to the user set value, the following transformation is required:

There is no nonlinear gain in the klystron phase, so it is necessary to remove the working phase information from the error term of the klystron output. Perform matrix operations as follows:

$$\begin{bmatrix} cos\varphi & sin\varphi \\ -sin\varphi & cos\varphi \end{bmatrix} * \begin{bmatrix} Acos\varphi - (A + \Delta A)cos(\varphi + \Delta\varphi) \\ Asin\varphi - (A + \Delta A)sin(\varphi + \Delta\varphi) \end{bmatrix}$$
$$= \begin{bmatrix} A - (A + \Delta A)cos\Delta\varphi \\ 0 - (A + \Delta A)sin\Delta\varphi \end{bmatrix} \quad (2)$$

Due to $\Delta\varphi \approx 0$, without considering the quadratic term of deviation, the calculation results are further simplified:

$$\begin{bmatrix} A - (A + \Delta A)cos\Delta\varphi \\ 0 - (A + \Delta A)sin\Delta\varphi \end{bmatrix} \approx \begin{bmatrix} -\Delta A \\ -Asin\Delta\varphi \end{bmatrix} \quad (3)$$

After removing the working phase, it can be seen from the Figure that to achieve the conversion from klystron output to input, only a change with *kgain* is required:

$$\begin{bmatrix} \frac{1}{kgain} & 0 \\ 0 & 1 \end{bmatrix} * \begin{bmatrix} -\Delta A \\ -Asin\Delta\varphi \end{bmatrix} = \begin{bmatrix} -\Delta A/kgain \\ -Asin\Delta\varphi \end{bmatrix} \quad (4)$$

Perform work phase compensation and obtain the compensation terms for klystron input as follows:

$$\begin{bmatrix} \cos\varphi & -\sin\varphi \\ \sin\varphi & \cos\varphi \end{bmatrix} * \begin{bmatrix} -\dfrac{\Delta A}{kgain} \\ -A\sin\Delta\varphi \end{bmatrix}$$

$$= \begin{bmatrix} \dfrac{-\Delta A\cos\varphi}{kgain} + A\sin\Delta\varphi\sin\varphi \\ \dfrac{-\Delta A\sin\varphi}{kgain} - A\sin\Delta\varphi\cos\varphi \end{bmatrix} \quad (5)$$

Compensate the compensation item to the original working point. Without considering the quadratic term of deviation, after further organizing the results, the user input target value for compensation can be obtained.

$$\begin{bmatrix} \dfrac{-\Delta A\cos\varphi}{kgain} + A\sin\Delta\varphi\sin\varphi \\ \dfrac{-\Delta A\sin\varphi}{kgain} - A\sin\Delta\varphi\cos\varphi \end{bmatrix} + \begin{bmatrix} A\cos\varphi \\ A\sin\varphi \end{bmatrix}$$

$$\approx \begin{bmatrix} \left(A - \dfrac{\Delta A}{kgain}\right)\cos(\varphi - \Delta\varphi) \\ \left(A - \dfrac{\Delta A}{kgain}\right)\sin(\varphi - \Delta\varphi) \end{bmatrix} \quad (6)$$

As shown in Figure 4, the user's target value after compensation is $\left(\left(A - \dfrac{\Delta A}{kgain}\right)\cos(\varphi - \Delta\varphi), \left(A - \dfrac{\Delta A}{kgain}\right)\sin(\varphi - \Delta\varphi)\right)$. After amplitude and phase hopping, the output value of klystron is $(A\cos\varphi, A\sin\varphi)$. So, for each pulse of microwave output from the klystron, the LLRF system first calculates the vector deviation between the initial measurement value within its pulse and the target. This deviation will be compensated for in LLRF excitation, thereby achieving suppression of the jitter after the pulse microwave.

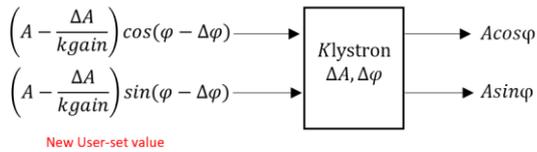

Figure 4: klystron output after feedforward compensation

### Key parameter calculation scheme

According to the analysis of klystron small-scale linear amplitude phase jump model and feedforward control algorithm, to suppress amplitude phase jump, it is necessary to measure the loop phase shift($\theta$), loop gain ($sgain$) and $kgain$ parameters in advance. Therefore, the design scheme is as follows:

Given that the target value set by the user is $(A\cos\varphi, A\sin\varphi)$. Considering the IQ bias measured by ADC, it is known that the relationship between the measured value($I_{meas}, Q_{meas}$) output by klystron and the target value is:

$$\begin{bmatrix} I_{meas} - I_{adcoffset} \\ Q_{meas} - Q_{adcoffset} \end{bmatrix}$$
$$= \begin{bmatrix} sgain*\cos\theta & -sgain*\sin\theta \\ sgain*\sin\theta' & sgain*\cos\theta \end{bmatrix}\begin{bmatrix} A\cos\varphi \\ A\sin\varphi \end{bmatrix} \quad (7)$$

After matrix operation, it can be concluded that:
$$\begin{bmatrix} A\cos\varphi \\ A\sin\varphi \end{bmatrix} = \begin{bmatrix} X & \cos\theta/sgain & \sin\theta/sgain \\ Y & -\sin\theta/sgain & \cos\theta'/sgain \end{bmatrix}\begin{bmatrix} 1 \\ I_{meas} \\ Q_{meas} \end{bmatrix},$$

$$\begin{bmatrix} X \\ Y \end{bmatrix} = \begin{bmatrix} sgain*\cos\theta & -sgain*\sin\theta \\ sgain*\sin\theta & sgain*\cos\theta \end{bmatrix}^{-1}\begin{bmatrix} -I_{adcoffset} \\ -Q_{adcoffset} \end{bmatrix} \quad (8)$$

By using multiple sets of target values and corresponding klystron output measurement values, the $sgain$ and $\theta$ parameters of the actual working point can be obtained. At the same time, based on the klystron input output model, the following matrix can be obtained:

$$[A_1 \quad A_2 \ldots A_n] = [Sgain \quad Sgain*kgain]$$
$$* \begin{bmatrix} A_i & A_i & \cdots & A_i \\ A_1 - A_i & A_2 - A_i & \cdots & A_n - A_i \end{bmatrix} \quad (9)$$

By fitting multiple sets of measured values and solving the matrix, the $kgain$ parameters near the user's actual working point can be obtained.

## EXPERIMENTAL ANALYSIS

Based on the intra pulse feedforward algorithm and implementation scheme, validation experiments were conducted during the pulse operation of Dalian Coherent Light Source (DCLS) linear accelerator to demonstrate the effectiveness and reliability of the algorithm.

### Automation application software

The implementation of algorithms and the automated measurement of key parameters require efficient software implementation. PyQt is a graphical interface development framework based on Python. The separation mechanism between the front-end interface and the back-end code effectively ensures the running efficiency of the program, making it easy for later maintenance and expansion. This article creates a user GUI application based on PyQt, which connects to the Epics system through the pyepics interface, as shown in Figure 5. The main functions of the software include: low-level excitation amplitude and phase adjustment, calculation of key parameters of the pulse feedforward algorithm, pulse feedforward switch, etc.

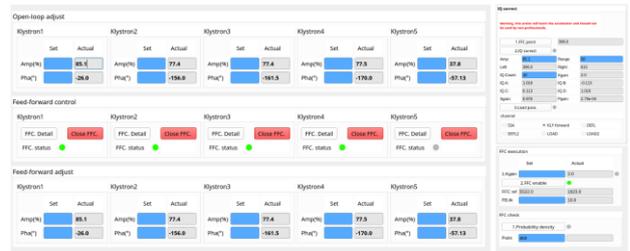

Figure 5: Intra-pulse feedforward algorithm application software

Firstly, the user determines the amplitude and phase operating point of low-level excitation through open-loop adjustment in the software. Then, according to the algorithm, key parameters near the amplitude and phase operating points in the klystron nonlinear model are calculated, including loop phase shift, *sgain*, and *kgain* parameters. This part is integrated into the software, and users only need to click on it. Finally, by measuring the output of klystron multiple times and taking the average value, the ideal measurement value of klystron without any jump in the klystron can be obtained. Together, they serve as the correction matrix for firmware calculation, and the firmware compensates in real-time based on the feedforward correction matrix. Merge and calculate all transformation matrices in the firmware, considering the delay of the entire physical loop from the DAC output to the ADC acquisition klystron output, the intra pulse feedforward will take effect at a position 700 ns after the intra pulse closed-loop point.

*Experimental verification*

The application verification of the intra-pulse feedforward algorithm and software was carried out in the DCLS deflection chamber. The RMS calculation results are shown in Figure 6. Taking into account the delay from the DAC output to the DAC acquisition klystron output, the intra-pulse feedforward will take effect with a delay of 700ns. Turn on intra-pulse feedforward at 4.1us. As can be seen from the Figure 6, it takes effect at 4.8us. This algorithm can effectively suppress the jitter between adjacent microwaves and improve the amplitude and phase stability (RMS) from 0.11%/0.2° to 0.1%/0.05°. This algorithm calculation is mainly used to suppress inter-pulse microwave layered jitter. It can be seen from the Figure 6 that the phase layered jitter phenomenon of microwaves is significantly stronger than the amplitude layered jitter. Therefore, the improvement of microwave phase stability by the algorithm is far better than the improvement of microwave amplitude stability.

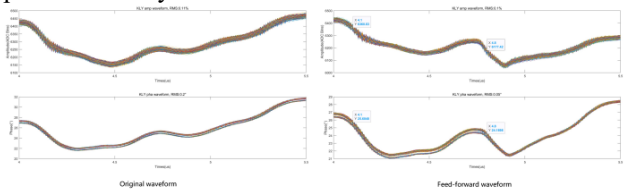

Figure 6: Feedforward algorithm experimental results

## CONCLUSION AND FUTURE WORK

This paper proposes an intra-pulse feedforward algorithm based on the klystron small-range nonlinear amplitude and phase jump model, develops user-level automated implementation software, and conducts experiments based on DCLS for verification. The results show that the algorithm can effectively suppress the jitter between adjacent microwaves and improve the amplitude and phase stability from 0.11%/0.2° to 0.10%/0.05°. In the future, we will further study the algorithm to improve the stability of low-level microwaves and accumulate theoretical and practical experience for the Shenzhen medium-energy high repetition rate X-ray free electron laser (S³FEL) device.